*Engineering and localization of quantum emitters in large hexagonal boron nitride layers*

Sumin Choi[1], Toan Trong Tran[1], Christopher ElBadawi[1], Charlene Lobo[1], Xuewen Wang[2], Saulius Juodkazis[2], Gediminas Seniutinas[1], Milos Toth[1,*] and Igor Aharonovich[1,*]

1. School of Mathematical and Physical Sciences, University of Technology Sydney, Ultimo NSW 2007, Australia
2. Centre for Micro-Photonics, Swinburne University of Technology, Hawthorn, Victoria , 3122, Australia

*Hexagonal boron nitride (hBN) is a wide bandgap van der Waals material that has recently emerged as promising platform for quantum photonics experiments. In this work we study the formation and localization of narrowband quantum emitters in large flakes (up to tens of microns wide) of hBN. The emitters can be activated in as-grown hBN by electron irradiation or high temperature annealing, and the emitter formation probability can be increased by ion implantation or focused laser irradiation of the as-grown material. Interestingly, we show that the emitters are always localized at edges of the flakes, unlike most luminescent point defects in 3D materials. Our results constitute an important step on the road map of deploying hBN in nanophotonics applications.*

In recent years layered materials, also known as van der Walls crystals, have attracted major attention across multiple fields of nanoscale science and technology[1, 2, 3, 4, 5]. For instance, transition metal carbides and nitrides (MXens) have been promising building blocks for energy storage and capacitors,[1, 6]. Similarly, transition metal dichalcogenides (TMDs) have been explored in nanoelectronics and nanophotonic applications mainly due to a unique transition from an indirect to a direct bandgap as their thickness is reduced to a single monolayer[2, 7, 8, 9, 10].

Hexagonal boron nitride (hBN) is a van der Walls crystal that has so far been used primarily as a capping or a separating layer for graphene and TMDs devices[4, 11, 12]. However, hBN has recently been shown to be the first known material that is naturally hyperbolic (meaning that the dielectric constants in the plane and out of plane have opposite signs), a property that has been leveraged to demonstrate sub-diffraction polariton propagation and sub-wavelength imaging with nanoscale resolution[13, 14]. hBN also possesses a wide bandgap of ~6 eV, that results in its ability to host many optically active defects over a wide spectral range[15, 16]. Research into isolated point defects of hBN has recently accelerated and several isolated defects have been characterized by scanning tunnelling microscopy[17] and optical confocal microscopy[18, 19, 20]. Indeed, one of the most fascinating properties of hBN is the ability to host ultra bright, room temperature single photon emitters that originate from localized defect states within the bandgap.

In this work we study the formation and localization of defects that act as single photon emitters in large (tens of microns wide) hBN layers. Ion implantation and laser processing are shown to enhance the formation probability of the defects. We study the photophysical properties of the emitters, showing their photoluminescence spectra, polarization properties, and photon emission statistics (autocorrelation functions). We use hBN flakes that are much larger than the spatial resolution of confocal photoluminescence microscopy and show that the emitters are always localized at boundaries or flake edges, in contrast to emitters in traditional 3D materials which are typically located away from surfaces and interfaces.

*Sample fabrication:* hBN layers were exfoliated from a bulk hBN material using standard scotch tape techniques. Figure 1a shows an optical image of the exfoliated material. Flakes with diameters of up to tens of microns were obtained. A reference sample and six substrates with exfoliated flakes were prepared for processing by ion implantation, laser ablation and electron irradiation. Ion implantation was explored using boron (B), boron-nitrogen (BN) complexes, silicon and oxygen ions. B and BN were selected to test whether the formation probability of intrinsic defects would increase, generating mostly vacancies and interstitials, while silicon and oxygen atoms were chosen to determine whether the emitters are related to common impurities. Indeed, oxygen is known to be trapped within hBN lattice during growth and exfoliation[21]. A summary of the sample processing details is shown in figure 1b. The laser ablation and electron beam irradiation treatments are described below. Unless noted otherwise, the presented data is from samples that were annealed for 30 minutes at $850^0$C in an argon environment[18, 22] either after exfoliation (reference sample) or after ion/laser processing.

*Defect creation using femtosecond laser pulse irradiation*

A Pharos laser system (Light Conversion Co. Ltd.) with a tunable pulse duration from $t_p$=230 fs to 10 ps, and an average power of 10 W, operated using a repetition rate of 200 kHz was used for laser processing of hBN. The second harmonic beam with a wavelength of 515 nm was focused onto the sample surface by an oil immersion objective with a NA of 1.4 (Olympus). The focus spot is around 450 nm (d=1.22λ/NA). The pulse energy was varied in 10% steps from 225 nJ to 90 nJ. A single pulse was delivered to each sample area, and the irradiated areas were separated by 5 μm. See supplementary information for an optical image of the fabricated array.

*Electron beam irradiation of exfoliated hBN flakes*

Bulk hBN was mechanically exfoliated onto a Si(111) substrate covered with a native oxide layer. The exfoliated flakes were rinsed with acetone and IPA and dried under flowing N2. The Samples were then loaded into a variable pressure FEI field emission gun scanning electron microscope. The system was pumped down to high vacuum and the chamber was filled with water vapour at a pressure of 8 Pa. The hBN flakes were then located using a magnetic field assisted gas ionization cascade detector, and electron beam irradiation was performed using a focused Gaussian electron beam which was scanned for one hour over an area of 600 μm$^2$. An accelerating voltage of 15 kV was used and the electron beam fluence delivered to the exposed area was 5×10$^{18}$ e-/cm2 (this is much lower than that needed to cause electron beam induced etching of hBN). The irradiation process is stopped by electron beam blanking.

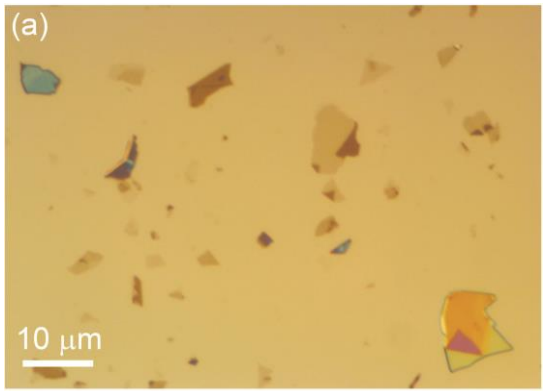

Figure 1: (a) Optical image of exfoliated hBN flakes. (b) Table of the investigated samples comprised of a reference sample, four samples that were implanted by ions, one processed by a laser beam and one by an electron beam.

| Sample | Irradiation | Annealing |
|---|---|---|
| B- irradiated | energy: 50 keV, fluence: 10$^{10}$ ion/cm$^2$, temp: RT | |
| BN- irradiated | energy: 50 keV, fluence: 10$^{10}$ ion/cm$^2$, temp: RT | 850°C 30 min 1 Torr Ar |
| O- irradiated | energy: 70 keV, fluence: 10$^{10}$ ion/cm$^2$, temp: RT | |
| Si- irradiated | energy: 50 keV, fluence: 10$^{10}$ ion/cm$^2$, temp: RT | |
| Reference | — | |
| Laser ablation | λ: 515 nm, pulse width: 230 fs; number of pulse: 1, temp: RT | |
| e-beam | energy: 15keV, e-fluence: 5 x 10$^{18}$ (e/cm2); temp: RT | — |

*Optical measurements*

Single photon emission characteristics were measured at room temperature using scanning confocal microscopy with a continuous-wave (cw) 532-nm laser (Gem 532™, Laser Quantum Ltd.). The laser was directed through a half wave plate and focused on the sample through a high numerical aperture objective lens (NA 0.9, Nikon). Scanning was performed using an X-Y piezo scanning mirror (FSM-300TM, Newport Corp.).The emission was collected using the same objective, filtered through a 532nm dichroic mirror and a long pass filter (Semrock), and coupled into a multimode fiber that served as a confocal aperture. Then a fiber splitter was used to split the light path to two avalanche photodiodes (APDs) (Excelitas Technologies™) for single photon counting and into a spectrometer (Acton SpectraPro™, Princeton Instrument Inc.). While the emission spectra were measured using the spectrometer, single photon detection was performed using a using a time-correlated single-photon-counting module (PicoHarp300™, PicoQuant™). Excitation polarization of the single photon emitters was controlled using a half wave plate, while the emission polarization was measured using a linear polarizer at maximum excitation polarization. The collected $g^{(2)}(\tau)$ curves were fit using a 3 level system equation

$$g^{(2)}(\tau) = 1 - (1 + a)\exp(-\lambda_1 \tau) + a\exp(-\lambda_2 \tau)$$

where $\lambda_1$ and $\lambda_2$ are decay rates for radiative and the metastable states, respectively. The polarization behaviour is fit by

$$I = I_0 \cos^2 \theta_i$$

where $I_0$ is the initial intensity and $\theta_i$ is the angle between the initial polarization direction of light and the transmission axis of the polarizer.

All optical characterization of the hBN emitters was performed at room temperature. Figure 2 (a-d) shows PL spectra recorded from samples implanted with B, BN, O and Si ions. The spectra in all cases are similar, showing a zero phonon line (ZPL) at ~ 600 nm and a weaker second peak near 650 nm. While some variation in the position of the ZPLs was observed, an absolute majority of the spectra exhibited two peaks. An investigation of 50 emitters revealed a similar range of spectra from each sample investigated in this work (ie: there are no statistically meaningful differences between the reference sample and samples processed by the various ion, laser and electron beam irradiation treatments). Interestingly, the difference between the two peaks seen in each spectrum is approximately 160 meV, indicating that all the emitters have similar structure within the hBN lattice[18, 20].

To ensure that the emission originates from localized single defects, we recorded the second order auto correlation function, $g^{(2)}(\tau)$, from each emitter. The functions are shown as insets in figure 2. The dips at zero delay time ($\tau=0$) confirm that the luminescence originates from single photon emitters. The data (blue dots) was fit using a standard three level model (solid black line)[18]. The deviations from zero at $\tau=0$ are attributed to background emissions.

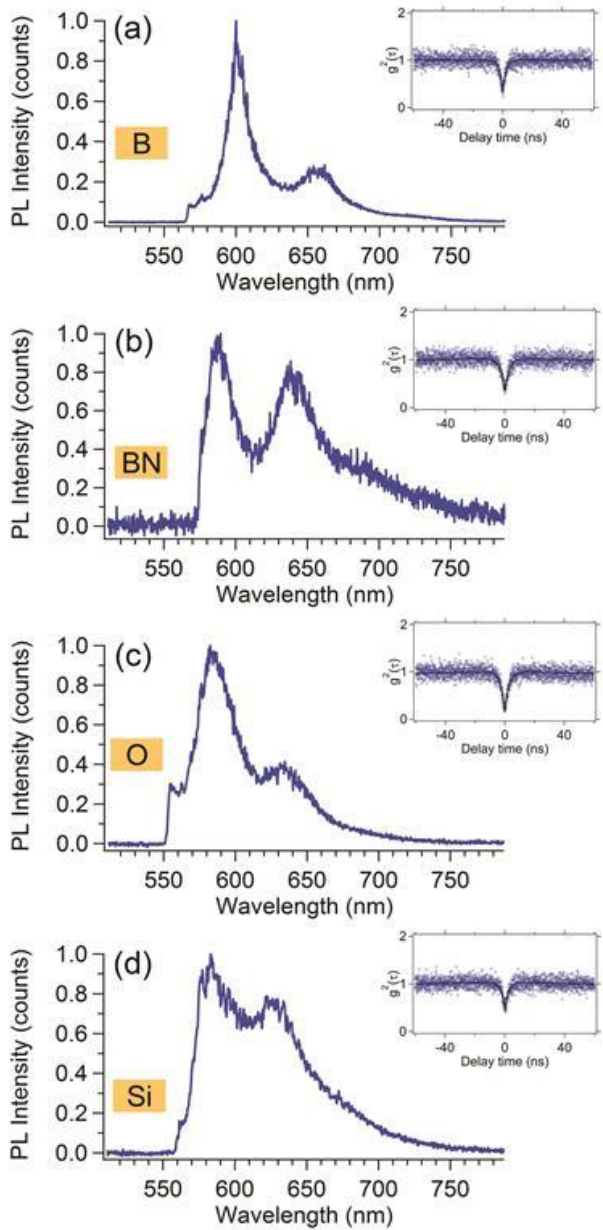

*Figure 2: (a-d) PL spectra from hBN flakes implanted with B, BN, O and Si ions. The insets are second order autocorrelation functions, $g^{(2)}(\tau)$, recorded from each sample, demonstrating the emitters are single photon sources. A spectrum from a reference sample and a corresponding $g^{(2)}(\tau)$ function are shown in the supplementary information (figure S1)*

Figure 3a shows a comparison of the number of single emitters found in the implanted flakes versus the reference sample (which has undergone the same annealing treatment used to activate the emitters,[18, 22] but without ion implantation). The implanted flakes show considerably more emitters than the reference. However, the ion species have little influence on the formation probability of the defects. This indicates that the main role of the bombarding ions is to introduce vacancies and to activate already present intrinsic point defects, rather than introduction of a foreign florescent defects (i.e. such as in the case of nitrogen implantation into diamond to produce nitrogen vacancy centers[23]). In addition, we observed that the emitters in the implanted flakes are mostly optically stable and do not exhibit blinking, while almost 50% of the emitters in the reference flakes show severe blinking and eventual bleaching. Figure 3 (b, c) shows typical intensity traces from a stable defect in an ion implanted flake and a blinking emitter in a reference sample. We therefore conclude that implantation can be used to increase the emitter activation probability as well as increase their photo-stability.

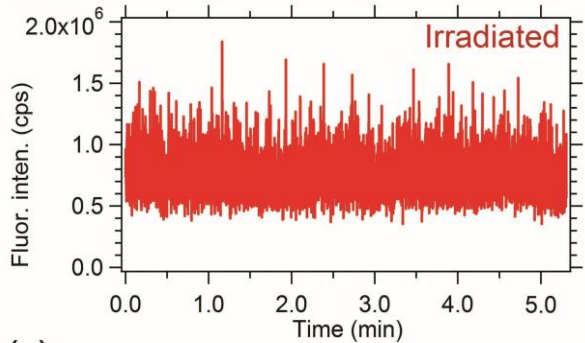
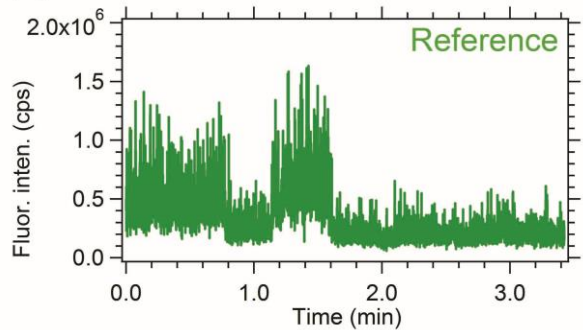

Figure 3. (a) A table comparing the number of formed emitters found in each ion-implanted sample and a reference sample that was only subjected to annealing. (b, c) Examples of stability curves from a single emitter in an ion implanted flake (b) and in a reference flake (c). Blinking followed by bleaching was much more common in the reference sample than in the implanted samples.

| Sample | Number of Emitters | Number of Flakes examined |
|---|---|---|
| B | 10 | 7 |
| BN | 10 (3 were blinking) | 12 |
| O | 10 | 12 |
| Si | 11 (3 were blinking) | 8 |
| Ref. | 7 (3 were blinking) | 16 |

Another important observation is the location of the emitters within the flakes. In contrast to prior work done using small (~ 200 nm wide) flakes of hBN, the samples used here are sufficiently large to resolve emitter locations within the flakes by confocal microscopy. Figure 4 shows three confocal maps from each implantation batch. White circles indicate quantum emitters. Remarkably, in all cases the emitters are localized at flake edges. While similar behaviour was reported for excitons in TMDs[24, 25, 26], such defect localization is an unexpected phenomenon for stable luminescent defects in semiconductors. Indeed, in 3D materials, the formation of stable, luminescent point defects near crystal edges and surfaces is extremely challenging, and the most stable and bright emitters are typically located in the bulk, deep within the crystal[27, 28, 29]. Emitter localization at interfaces is desirable for device fabrication as it can improve ultimate control over emitter placement and coupling to photonic and plasmonic cavities. We note that due to the finite resolution of the confocal microscope, which is ~ 300 nm in our experimental setup, we cannot conclusively say how far the emitters are from the flake edges. However, there is no compelling reason to suggest that point defects are localized tens or hundreds of nanometers away from an edge, and it is therefore most likely that the emitters decorate the edges and crystal boundaries. Interestingly, extended line defects in BN with different chemical terminations and geometrical variations have been modelled and predicted to have unique optoelectronic properties that can result in confined, optically active systems[30]. Detailed atomistic modelling and super resolution imaging will be required to elucidate this behaviour further.

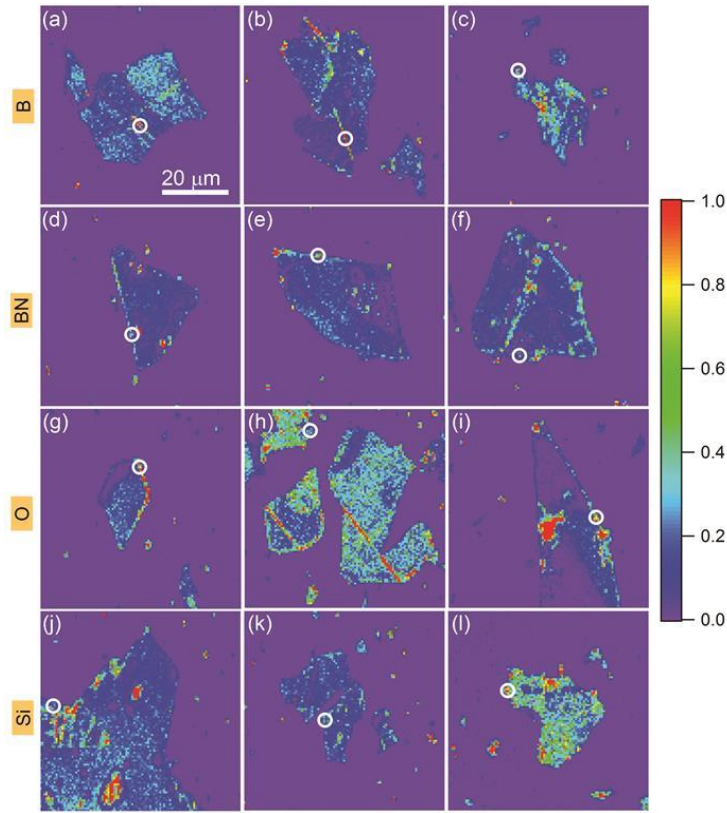

*Figure 4: Confocal maps from (a-c) B, (d-f) BN, (g-i) O and (j-l) Si implanted hBN flakes, demonstrating unambiguously that the emitters are always localized at flake edges. Large bright features seen away from flake edges, as in map (i), do not exhibit photon antibunching, and do not posses the spectral characteristics of the single photon emitters discussed in this study.*

figure 5 (a-d) show examples of excitation and emission polarization recorded for different emitters from different implantation batches. More examples are shown in the SI. While most of the emitters are fully polarized in both excitation and emission, we did observe numerous emitters that did not show full extinction. This is most likely because the flakes were not properly adhered to the substrate post exfoliation, creating an angle between the flake and the excitation laser beam. The polarization behaviour is therefore indicative of a dipole like emitter, with a fast polarization axis, in accord with the earlier studies[18, 20]. The misalignment between the excitation and the emission polarizations is likely due to redistribution of the excited electronic states.

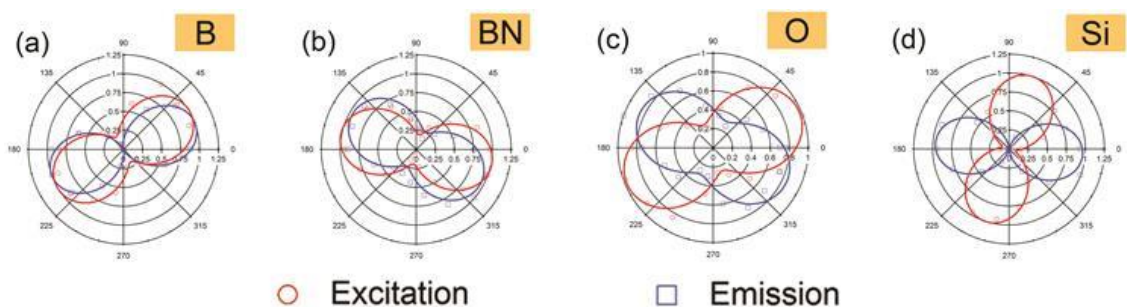

*Figure 5: (a-d) Example of excitation (red circles) and emission (blue squares) polarization plots of single emitters from each of the implanted samples. All the emitters exhibit dipole like behaviour in both excitation and emission.*

Next, we characterize flakes that were processed using an ultrafast pulsed laser operated at a power just below the threshold for rapid ablation[31]. Ultra-short laser pulses are efficient sources of free electron acceleration due to high peak intensities. Free electrons accelerated to energies larger than the bandgap are efficient in defect formation and chemical bond breaking. Colour center formation in dielectric materials is typical under fs-laser irradiation at such fluence/irradiance[32]. Figure 6(a) shows a confocal map of the sample, while figure 6(b) shows a photoluminescence spectrum recorded from a single emitter found in these flakes. The inset is a corresponding $g^{(2)}(\tau)$ curve that confirms single photon emission from this defect. Similarly to the ion implantation case, the emitter exhibits full polarization behaviour in both excitation and emission as is seen in figure 6(c)

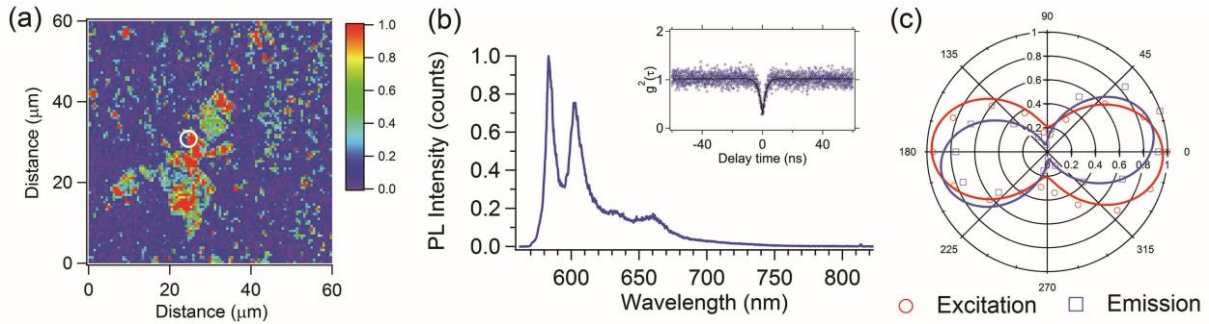

*Figure 6. Fabrication of emitters using laser processing. (a) Confocal map of the hBN flakes. The white circle indicates the presence of the location of a single photon emitter. (b) PL spectrum recorded from the emitter. Inset: $g^{(2)}(\tau)$ curve confirming it is indeed a single photon emitter. (c) excitation and emission polarization curves (red circles and blue squares, respectively,) recorded from this emitter.*

The original size of the flakes was large, similarly to the ones shown in figure 1a. However, the laser process caused some of the flakes to break up into many small fragments. This resulted in more observed emitters per similar scan area.

Finally, we use a deterministic electron beam induced irradiation technique to fabricate the emitters in specific hBN flakes (this technique has been used previously to fabricate emitters in sub-micron hBN flakes that were too small to determine emitter locations within the limits of diffraction-limited confocal microscopy[22]). Figure 7 (a, b) shows a flake before and after electron beam irradiation by a 15 keV electron beam in a $H_2O$ vapour environment (see methods). Figure 7(c) shows photoluminescence spectra recorded from the pristine flake (blue) and after the irradiation treatment (red). Note that the spectra were recorded from the same location, as indicated by the white circles in (a) and (b). The inset is a $g^{(2)}(\square\square)$ function that proves the probed emitter is a single photon source. Figure 7(d) shows the corresponding polarization measurements from the same emitter. This sample was not annealed after electron beam irradiation because the annealing is not required for emitter activation, in contrast to emitters generated by the ion and laser irradiation treatments. This can be explained by the fact that ion implantation and laser irradiation generates significant damage in the hBN lattice that partly recovers during annealing. On the other hand, irradiation by 15 keV electrons in $H_2O$ vapour is a more subtle process that chemically reforms the lattice, with minimal damage to the surrounding crystallographic environment. The electron beam approach is therefore appealing as it allows emitter fabrication and localization in a single step, without the use of lithographic masks or post-processing treatments. The emitters fabricated by an electron beam had the same spectral and polarization characteristics as those made by ion and laser irradiation, and they were located consistently at flake edges. Further work is needed to determine whether the electron beam creates new defects or activates pre-existing defects that are present in as-grow hBN.

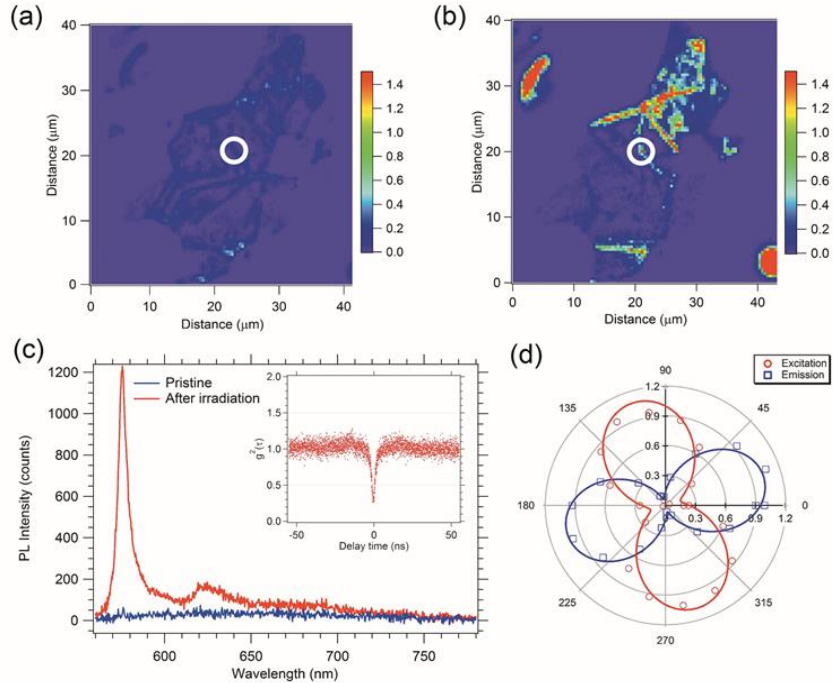

*Figure 7. Fabrication of emitters by electron beam irradiation. Confocal map of the same flake before (a) and after (b) electron beam irradiation. The energy of the beam is 15 keV. (c) Spectra recorded from a particular location before (blue curve) and after (red curve) irradiation. Inset, a $g^{(2)}(\tau)$ curve confirming the formed defect is a single photon source. (d) excitation and emission polarization from the same defect. The sample was not annealed after electron irradiation.*

To summarize, we presented a comprehensive study of single emitters in large layered hBN. We find that ion implantation, laser ablation and annealing are efficient methods to generate the emitters, however, the implantation species have little influence. We also observe that the emitters are always localized at flake edges, indicating that a flake morphological defects may play a role in the formation of these quantum emitters. Finally, we showed that electron beam irradiation can be used to fabricate the emitters in a particular flake. Overall, the emitters are polarized and optically photostable and therefore are very promising for future quantum photonics and quantum optotoelectronic applications. Further studies are required to unveil the exact structure of the single emitters large layered hBN.

**Acknowledgments**

Ion implantation was performed in the Department of Electronic Materials Engineering, RSPE (ANU). The work was supported in part by the Australian Research Council (ARC) (DP140102721, IH150100028 ARC Research Hub for Integrated Device for End-user Analysis at Low-levels), FEI Company and by the AOARD grant FA2386-15-1-4044. I. A. is the recipient of an ARC Discovery Early Career Research Award (DE130100592).